\documentstyle[aps,epsfig,multicol]{revtex}

\newcommand{\figwidth}{0.9\linewidth}
\newcommand{\insetwidth}{0.32\linewidth}

\begin{document}

\title{Mapping a depinning transition to polynuclear growth}

\author{G.J. Szab\'o$^{1,2}$ and M.J. Alava$^{1}$}

\address{$^1$Helsinki University of Technology, Laboratory of
Physics, P.O.Box 1100, FIN-02015 HUT, Finland}
\address{$^2$Technical University of Budapest, Department of Theoretical
Physics, H-1111 Hungary}

\maketitle
\begin{abstract}
\noindent
We develop a phenomenological mapping between submonolayer polynuclear
growth (PNG) and the interface dynamics at and below the depinning
transition in the Kardar--Parisi--Zhang equation for a negative
non-linearity $\lambda$. This is possible since the phase transition
is of first-order, with no diverging correlation length as the
transition is approached from below. The morphology of the
still-active and pinned configurations and the interface velocity are
compared to the PNG picture. The interface mean height scales as
${\mathrm{erf}}(t)$.

\medskip
\noindent {\it PACS \# \ 05.70.Np, 75.50.Lk, 68.35.Ct, 64.60.Ht}
\end{abstract}

\begin{multicols}{2}[]

\section{Introduction}
The dynamics of driven manifolds in random media presents many
examples of non-equilibrium phase transitions. The interest lies often
in how an object interacts with a quenched disorder environment. If
the driving force is small enough and the temperature zero, the
manifold, e.g. a domain wall in a magnet, or an interface between two
phases, can get pinned. This means, quite simply, that its velocity
becomes zero. Therefore one can discuss the physics in terms of an
order parameter (velocity) and a control parameter (external force).
At and close to the critical force value $F_c$ the driven interface
can develop critical correlations due to the frustration that arises
from the competition between, typically, elastic forces and the
randomness of the medium
\cite{Barabasi,Hah95,Kardar,Fisher,driven_depinning,sneppen}.

An example is provided by the quenched Kardar--Parisi--Zhang (QKPZ)
equation,
\begin{eqnarray}
{\partial h({\mathbf x}, t) \over \partial t} = \nu \nabla^2 h +
{\lambda \over 2} (\nabla h)^2 + F + \eta({\mathbf x}, h).
\label{EKPZ}
\end{eqnarray}
where $h$ is the interface height, $\nu$ a surface tension, $\lambda$
the strength of the nonlinearity characterizing the KPZ class
\cite{KPZ}, $F$ a driving force, and $\eta$ is the quenched noise. If
$\lambda$ is positive and stays finite in the limit $F \rightarrow
F_c^+$, the depinning phase transition in 1+1 dimensions can be mapped
to directed percolation (DP) via a geometrical construction. In this
way one can compute the critical exponents through the knowledge
concerning the DP exponents, and the particular model where this is
explicitly possible has been coined 'directed percolation depinning'
one for obvious reasons \cite{DPD,tang_leschhorn,family_vicsek}.

If $\lambda$ is negative, however, the character of the transition
changes completely. Fig. 1 shows an example of an interface at $F \sim
F_c$. The depinning in this case was investigated by Jeong, Kahng, and
Kim~\cite{korean}, who showed that the transition becomes of a {\em
first-order} kind and that this persists even in the presence of
finite tilts below a critical $m_c$. The aim of this article is to use
a new geometrical construction to discuss the morphology in the case
depicted in Fig.~1, similarly in spirit to the DPD-mapping of the
$\lambda>0$-case.

The essential physics in the $\lambda<0$ QKPZ class, at and below
$F_c$ arises from ``facet formation'', the fact that the slopes of the
triangular mounds are just dependent on the parameters of the
equation, and from ``valley formation''. A look at Fig. 1 where the
interface is shown at various times $t_i$ implies that it is these two
mechanisms that are responsible for the final shape of an interface
that gets pinned. In this article we map the two-dimensional growth of
the QPKZ interfaces to a one-dimensional sub-monolayer polynuclear
growth model (PNG) \cite{frank} based on this picture of a
statistically constant growth velocity of the facets, and random
pinning events on the roughly speaking planar areas of growth. This
implies that one can understand the pinning of an interface in a
quenched environment by considering a model with 'thermal noise',
only. The reason why this is possible is the first-order character of
the transition: there is no diverging correlation length, and
therefore the physics of the depinning can be understood based on a
local picture of events.  Notice that the PNG in the multilayer
version is actually in the thermal KPZ-class, and that the relation in
our case between a growth model and the PNG is indirect.

\begin{figure}
\begin{center}
\includegraphics[width=\figwidth]{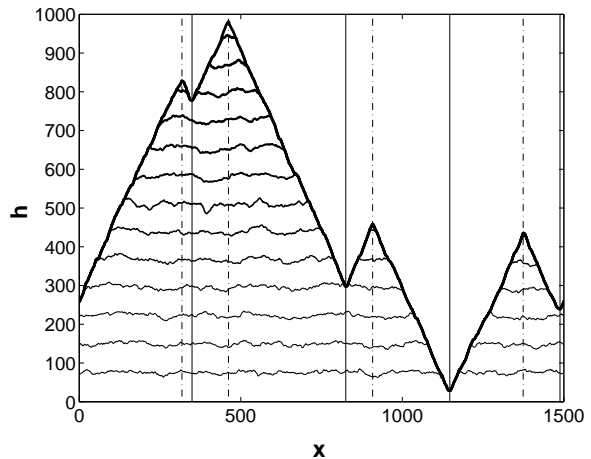}
\end{center}
\smallskip
\caption{Temporal evolution of a QKPZ interface at an external driving
of $F=14.5$. The snapshots have been taken every 1000th time step. The
vertical slicing lines designate the positions of ``valleys'' and
``peaks'' located by the algorithm described in the text.}
\label{snapshot}
\end{figure}

Section two of this paper introduces the actual numerical QKPZ model
and the relevant parameters. In Section three, we discuss in detail
the mapping between the QKPZ and the PNG models including a
measurement of the model parameters for use in the PNG model. We also
discuss the relevant parts of the PNG literature. In Section four the
free parameters (growth velocity, pinning probability per unit time)
are discussed and computed, and a basic comparison of the QKPZ
interfaces with those that come from the PNG representation is
presented. Section five finishes the paper with a discussion.

\section{Numerics}

The model we focus on is a $1+1$ dimensional QKPZ system with the
spatial coordinate $x$ and time $t$ discretized, while the interface
height $h$ is a continuous number. We use periodic boundary conditions
with flat initial configuration. The forthcoming results have been
obtained from systems in which the equation coefficients are
invariably $\nu=10$, $\lambda=-4$, and $g=20$.

The system is updated simultaneously for all lattice sites in every
time step according to the finite differences version of the governing
equation~(\ref{EKPZ}) with $\Delta t=0.01$:
\begin{eqnarray}
f_i(t) = \nu[h_{i+1}(t) + h_{i-1}(t) - 2h_i(t)] + {}
\nonumber\\
{}+{\lambda \over 8}[h_{i+1}(t)-h_{i-1}(t)]^2+F+g \eta_i([h_i(t)])
\nonumber\\
h_i(t+\Delta t) = \left\{ \begin{array}{ll}
h_i(t) + f_i \Delta t & \textrm{if $f_i(t)>0$}\\
h_i(t)                & \textrm{otherwise}
\end{array} \right.
\label{Esim_1d_cont_lesch}
\end{eqnarray}

The value of the noise term $\eta_i$ at integer positions of the
height field denoted by $[\cdot]$ is chosen randomly from the interval
$[-1; 1]$ with a uniform distribution and is unchanged in time. The
choice of the noise term is meant to mimic the simulation carried out
in~\cite{korean}. Nevertheless, the particular form of $\eta_i$ has a
mere quantitative influence on some measurables such as the critical
driving force $F_c$; the qualitative behavior of the systems have been
checked to yield consistent results independently of the various
options for the representation of $\eta_i$.

We study the morphology of slightly subcritical interfaces with the
mapping to the PNG model in mind. To this end we identify the
macroscopic facets along the interface locating the ``valleys'' and
``peaks'' in the entirely evolved and stopped systems.

The algorithm to find the extrema is composed as follows. As a first
step the pinned interface is scanned for all of the local minima and
maxima positions. Subsequently after that, the pair of adjacent
minimum/maximum with the smallest distance to each other is removed
and this step repeated until the distance between the two closest
extremal positions is no more than a specific parameter. Applying this
procedure we eliminate the noise-like small height fluctuations of the
interface. The resolution of the facets depends on the actual choice
of the smallest distance parameter and we use a few lattice units for
that.

\section{Mapping to the polynuclear growth model}
\subsection{The interface growth in one dimension}
Following the time evolution of some interface configurations like the
one in Fig.~\ref{snapshot}, it suggests itself that the pinning events
resulting in the valleys of the final morphology take place in a
seemingly random way on the growing terraces. The interface is
observed to propagate with a uniform average vertical velocity at the
plateaus of the facets, in other words, the height of the separated
terraces of the interface is on the average on the same level at all
times.  The straight sides of the facets mean that the horizontal
distance between the end points of neighboring facets at the level of
propagation grows in a linear fashion with time, following to first
order from a balance between the nonlinearity and the driving force in
Eq.~(\ref{EKPZ}).

If we consider the \emph{h} axis of Fig.~\ref{snapshot} as time and
the sides of the final interface facets as phase boundaries in this
model, we can outline a picture of a model mapping as follows. We
define a one dimensional lattice of cells on the grounds of the PNG
model with two-state filling possibilities: a site is either occupied
by a particle or empty, with an initially empty configuration. During
the simultaneous updating steps of the lattice, a seed particle may be
deposited to any of the empty cells with a certain probability in
every time step. The actual deposition probability and place may
depend on various factors which will be discussed later.

\begin{figure}
\begin{center}
\includegraphics[width=\figwidth]{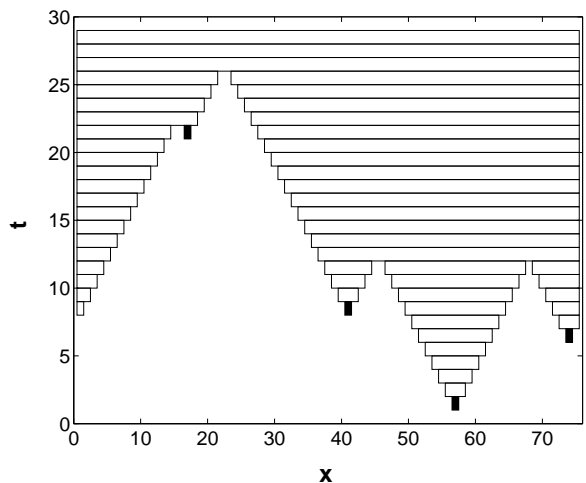}
\end{center}
\caption{Time evolution of PNG islands with the seeds filled
black. The mapping is performed for the interface configuration
depicted in Fig.~\ref{snapshot} with a system size rescaling factor of
$\Lambda = 0.05$.}
\label{png_mapping}
\end{figure}

Deposition of seeds creates growing islands with one particle added
per time step at both ends according to the formal rule of
\begin{eqnarray}
s_i(t+1) = \left\{ \begin{array}{ll}
1 & \textrm{if $s_{i-1}(t)=1$ or $s_{i+1}(t)=1$}\\
s_i(t) & \textrm{otherwise}
\end{array} \right.
\label{Efilling_rule}
\end{eqnarray}

$s_i(t)$ denotes the filling state of the cell $i$ at time $t$. An
occupied site is marked by 1 while an empty one is marked by 0. In
other words (\ref{Efilling_rule}) means that a cell gets filled if
either of its neighbors is already filled. The system has periodic
boundary conditions so $s_{i-1}=s_L$ for $i=1$ and $s_{i+1}=s_1$ for
$i=L$. $L$ is the lattice size and the indices run from $1$ to $L$.

It is obvious that if we associate the empty cells in this model with
the still-growing parts of the interface and the filled cells with the
uncovered regions, the time evolution of the system would give a very
similar picture to the final configuration of Fig.~\ref{snapshot}, if
the way the seeds are deposited is chosen the right way. An example of
the mapping is demonstrated by Fig.~\ref{png_mapping}.

\begin{figure}
\begin{center}
\includegraphics[width=\figwidth]{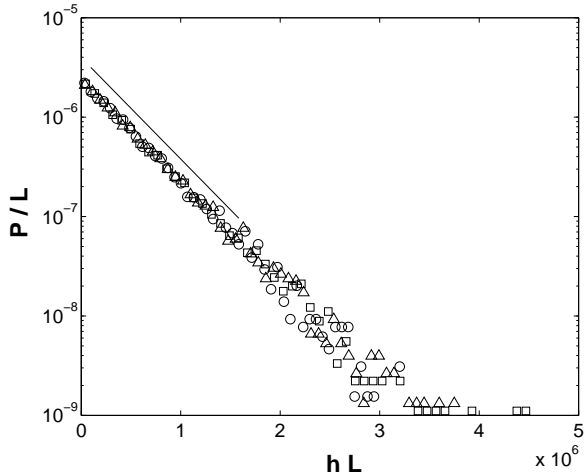}
\end{center}
\caption{The probability distribution $P$ of the first pinning height
$h$ in the KPZ system, rescaled by the system size of 500, 1000, and
2500 for {\Large $\circ$}, $\triangle$, and $\Box$, respectively. The
driving force is $F=14.8$. The averaging is done for 10000 interface
configurations in each case. The rescaled distribution is found to
decay exponentially with an associated exponent of $b \approx 2.35
\cdot 10^{-6}$ in the range of the parallel line shown in the figure.}
\label{first_pinning}
\end{figure}

Fig.~\ref{first_pinning} suggests that the probability the first seed
is introduced at a given time should decay exponentially in time,
which is the earmark of a geometric process. This means that the
probability of a seed deposition (the creation of the first pinned
site) has a constant probability per each time step. If this is true,
then the growth closely resembles the submonolayer polynuclear growth
model (PNG), or the Kolmogorov--Avrami--Johnson--Mehl (KAJM)
model~\cite{KAJM1,KAJM2,KAJM3}. (The PNG model is originally defined
so that the growth itself takes place in many layers atop each other.)

\begin{figure}
\begin{minipage}[t]{\linewidth}
\begin{center}
\includegraphics[width=\figwidth]{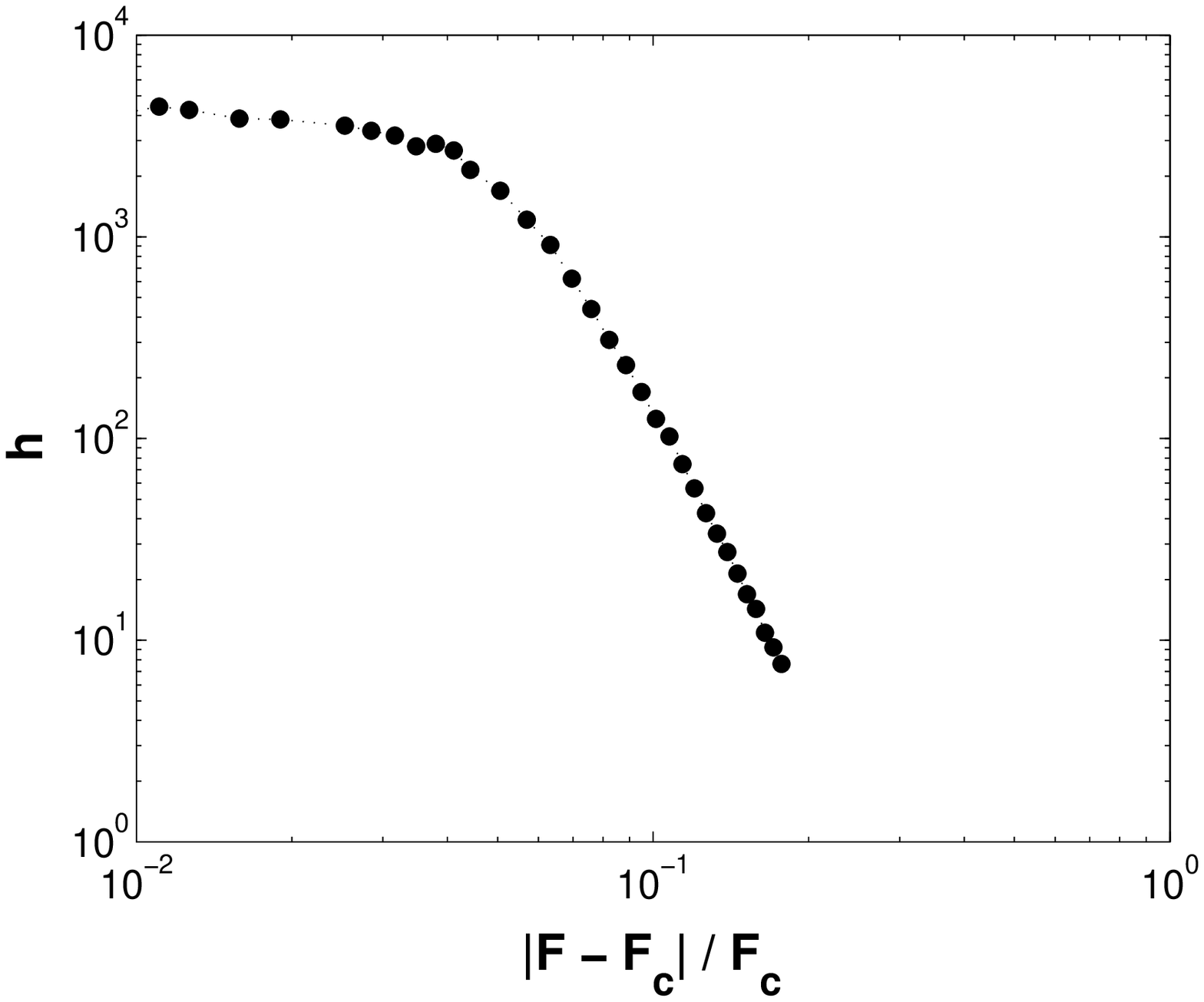}
\end{center}
\vspace{-2.5in}
\hspace{1.9in}
\includegraphics[width=\insetwidth]{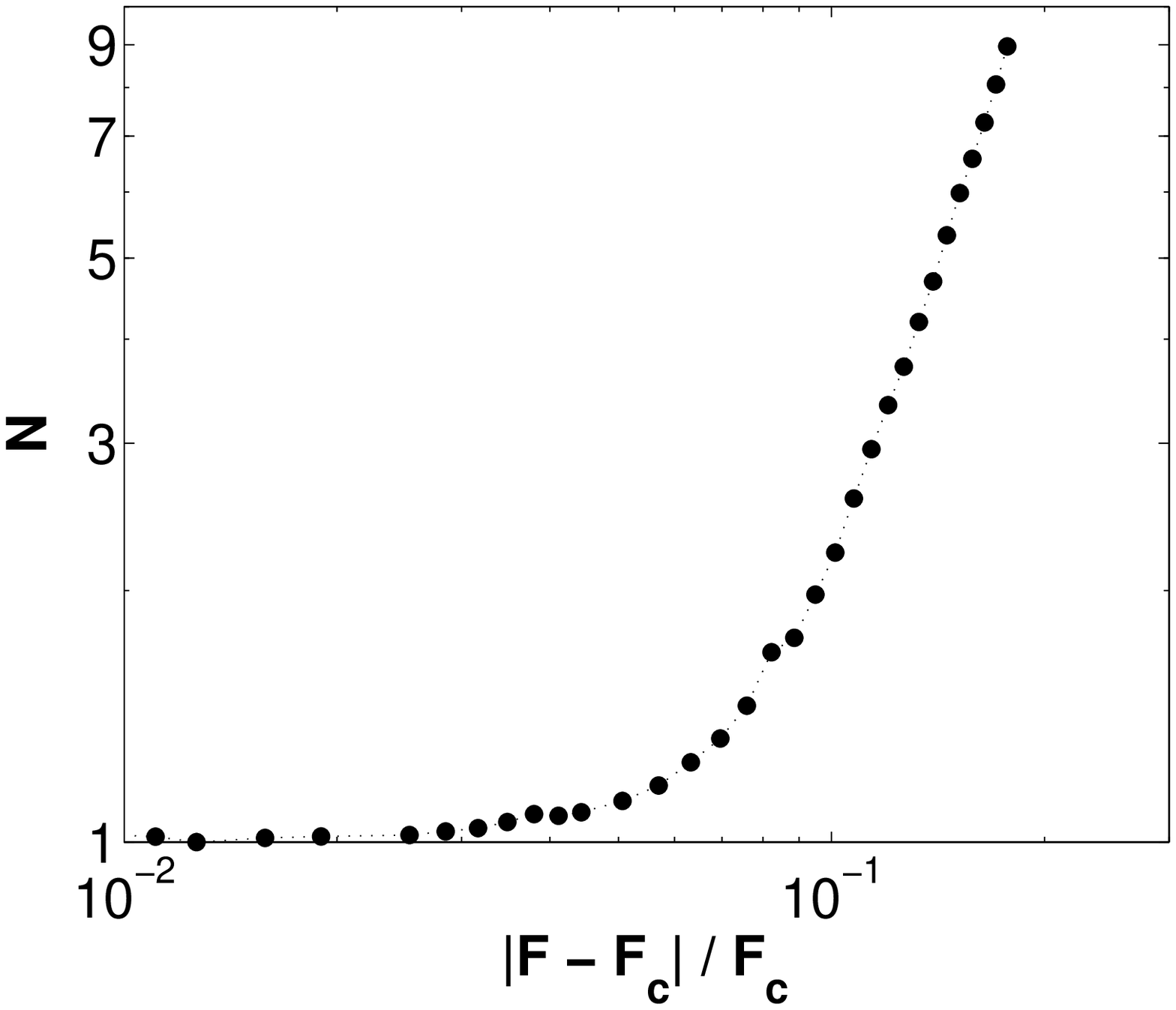}
\end{minipage}
\caption{The first pinning height $h$ as function of the driving force
$|F-F_c|/F_c$ for a system of 500 lattice units. The saturation is
caused by limitations imposed on the maximum time configurations are
allowed to evolve because of practical reasons. The inset shows the
number of pinnings taking place.}
\label{force_function}
\end{figure}

Another crucial question is: how is the pinning probability or the
nucleation probability in the PNG picture dependent on $F$?
Fig.~\ref{force_function} illustrates the progress of two quantities
close to the estimated $F_c$. Both the average number of pinning
events and the average height (time) of the first pinning event scale
roughly as power laws with $|F-F_c|$. Since the depinning transition
is of first order, it is not a priori obvious how these quantities
should behave as $F \rightarrow F_c^-$. The most natural guess is that
the pinning probability per site per time-step goes to zero
continuously, which would indicate that $h(F)$ and $N(F)$, as depicted
in Fig.~\ref{force_function}, should approach infinity and zero,
respectively with cut-offs imposed by the maximum time the interfaces
are followed, and the system size, respectively.

\subsection{Interface morphology}
The valleys and peaks are identified like described in Section II.
The side gradients of the facets are measured on both sides and are
displayed for various system configurations in
Table~\ref{Tmorphology_res}. The height and position of the pinning
events are also determined and recorded.  $F$ is chosen so that the
system develops only a limited number of facets; this way, the
uncertainty of facet identification and measurement stays under
control. On the other hand, we are mainly interested in the behavior
when $F \lesssim F_c$. We do not determine $F_c$ with very great
accuracy; good estimates can be found in~\cite{korean}.

According to Table~\ref{Tmorphology_res}, the facet side gradients are
independent of system size for a given $F$. 
Additionally, the left and right hand side slopes are the same
as well, as expected in the absence of a global tilt. 
In contrast to this, the number of pinning events taking place
in the systems is subject to finite size scaling. The error values
associated with the averages are apparently large (30--40\%).

\subsection{PNG theory}
For the next section where the growth statistics are compared with the
one-dimensional picture some results are needed about the interesting
quantities. In the ordinary KAJM theory islands are nucleated on a
line randomly, with a rate $\gamma$ and grow at a constant velocity.
To map to the KPZ case one needs a function that relates that velocity
to the KPZ mound slopes as discussed below. If one has $N(t)$
nucleated islands (here pinned parts of the interface) then the
fraction of uncovered space (here growing fraction of interface)
$S(t)$ satisfies the rate equation $dS/dt = - N$ with a
proportionality factor chosen as unity.  It is easy to understand that
in a large system the temporal fluctuations of the slopes of the
growing KPZ interface average out, to an average growth velocity in
the KAJM picture. This holds if the slopes do not depend on e.g. the
size of the terrace they are adjacent to which seems trivially
true. Then the only factor that needs to be checked is whether the
nucleation events happen randomly, with a constant density as in the
simple KAJM as discussed above.

In the one-dimensional picture, one can proceed by defining size
distributions for the islands, for the gaps between the island, and
finally the density of islands that contain a fixed number of seeds,
$n$. The last one implies that such islands result from the
coalescence of $n$ single-seed islands. Such distributions describe
exactly the KPZ interface in the KAJM language (assuming that the
mapping to PNG/KAJM works out).  The last of these distributions is
hard to compute since it involves a solution of a Smoluchowski-like
equation for the densities for different $n$. We quote here some of
the appropriate results from~\cite{bennaim1,bennaim2}. First, the
island number density,
\begin{equation}
N(t) \sim t e^{-t^2},
\label{n1}
\end{equation}
corresponds to the number of valleys in the KPZ picture 
(as, also, a function of average interface height, $N(h)$).
Second, the uncovered fraction gives directly the interface
velocity after a proper normalization, and reads
\begin{equation}
S(t) = e^{-t^2/2},
\label{n2}
\end{equation}
implying a fast decay of the growth. 

\section{Comparison of the PNG picture with the KPZ model}

Following the considerations above, let us denote the probability of a
seed deposition per lattice unit in the PNG model $\Pi$ for every time
step. $\Pi$ should be tuned appropriately so that the average number
of seed depositions equals the average number of pinning events or
valleys in the KPZ system. This is the only parameter that affects the
``morphology'' in this system, while another one, an artificial ``side
gradient'' $\mu$ has also to be introduced to transform the time scale
of the PNG model to length scale in the KPZ counterpart.

Notice that the virtual valleys and summits from the PNG model after
the scale transformation are not rounded.  The presence of the surface
tension term in the KPZ equation counteracts such sharp corners,
though. Still, if the strength of the surface tension coefficient
$\nu$ is sufficiently small, the resulting interface is very similar
to the one obtained from the PNG model.

To achieve proper quantitative predictions and match between the KPZ
morphology statistics and the PNG picture, we need to tune $\Pi$ and
$\mu$ so that measuring the first pinning height (seed deposition
time) distribution, we end up with the same decay exponents. We relate
the general height in the KPZ model $h_{\mathit{KPZ}}$ to time in the
PNG system $t_{\mathit{PNG}}$ as $h_{\mathit{KPZ}} = \mu
t_{\mathit{PNG}}$. The probability of the first seed deposition
follows a geometric process:
\begin{eqnarray}
P(t_{\mathit{PNG}}=\tau) = P(h_{\mathit{KPZ}}/\mu=\tau) =
(1-p)^{\tau-1} p=
\nonumber\\
=(1-p)^{\chi_{\mathit{KPZ}}/\mu-1} p
\nonumber\\
\log P(t_{\mathit{PNG}}=\tau) = \log {p \over {1-p}}+{{\log (1-p)}
\over \mu} \chi_{\mathit{KPZ}},
\label{geometric}
\end{eqnarray}
if $p$ denotes the seed depositing probability for the entire length
of the PNG system, that is, $p=\Pi L_{\mathit{PNG}}$.
$\chi_{\mathit{KPZ}}$ is the first pinning height (the altitude of the
lowest valley) in the corresponding KPZ system. Comparing
Fig.~\ref{first_pinning} to Eq.~(\ref{geometric}) we find that the
assumption of $p=\mathit{const.}$ is valid.

The experimental value of the probability decay exponent $\beta$ in
the KPZ system should thus equal $\log (1-p)/\mu$ expressed by the PNG
parameters. The Taylor series expansion of the later results in
$p=-\beta \mu \ln 10$ up to first order.

If the horizontal scale transformation between the KPZ and PNG systems
is given by $L_{\mathit{PNG}}=\Lambda L_{\mathit{KPZ}}$, then $\mu$
can simply be calculated from the measured facet side gradients in the
KPZ system, summarized in Table~\ref{Tmorphology_res}. Then if $s$
denotes the empirical value of the side slope,
$\mu=s/\Lambda$. According to these, the normalized seed deposition
probability $\Pi$ is obtained as
\begin{eqnarray}
\Pi=-{{\beta s \ln 10} \over {\Lambda L_{PNG}}}.
\end{eqnarray}

The discussion above assumed that the parameter matching is performed
between two particular lattices of specific features, most importantly
the number of cells in the systems. Considering the data collapse of
Fig.~\ref{first_pinning}, though, an additional relation for $\beta$
emerges instantly: $\beta \sim L_{\mathit{KPZ}}$, provided that the
equation parameters ($\nu$, $\lambda$, $F$, $g$) are otherwise
fixed. Since $s$ does not explicitly depend on $L_{\mathit{KPZ}}$, and
if $\beta=b L_{\mathit{KPZ}}$,
\begin{eqnarray}
\Pi=-{{b L_{\mathit{KPZ}} s \ln 10} \over
{\Lambda (\Lambda L_{\mathit{KPZ}})}}=
-{{b s \ln 10} \over {\Lambda^2}}.
\end{eqnarray}

A further consequence of the fact that $P \sim \Phi(h
L_{\mathit{KPZ}})$ for the first pinning probability with a suitable
scaling function $\Phi$, is that the general pinning probability per
time unit in propagating fronts of a QKPZ system is {\em linearly
proportional} to the integrated local interface velocity, $v(x,t)$,
\begin{eqnarray}
P_{pinning}(t) \sim \int v(x, t) \, dx.
\end{eqnarray}

Using the parameters above it is possible to completely recover the
exact scaling law of the first pinning event in a system conforming to
the KPZ equation. It may still be
uncertain, though, whether the subsequent pinning events follow the
same governing rules as the first one, in particular the
assumption that the seed creation probability is
linearly proportional to the island length.

\begin{figure}
\begin{center}
\includegraphics[width=\figwidth]{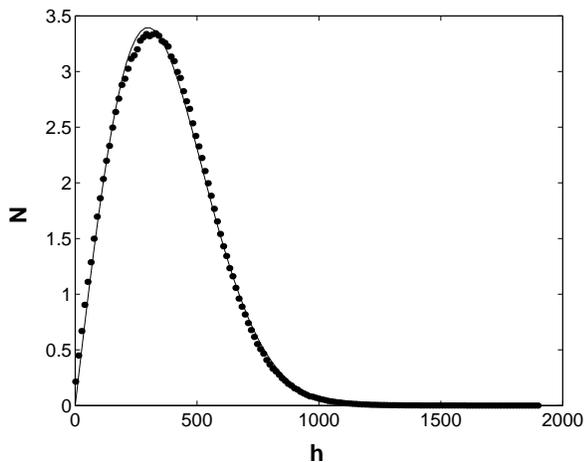}
\end{center}
\caption{The ``island number'' distribution $N(h)$ calculated for a
KPZ system according to the text. Since the interface propagation
velocity is constant, $h$ can be appropriately rescaled so as to yield
time in a PNG system. The system size is 2500, while the driving force
is $F=14.5$. The number of averaging iterations is $10000$. The
theoretical function of $N(h)=A \, h \, exp(-B h^2)$ is fitted to the
empirical data points minimizing the quadratic error and giving the
results of $A=5.84 \cdot 10^{-3}$ and $B=2.33 \cdot 10^{-6}$. The
slight shift of the distribution maximum may come from the fact that
an initially flat interface cannot get pinned before the the necessary
fluctuations develop. On the other hand, the omission of very small
facets can have an unwanted effect on the data as well.}
\label{island_number}
\end{figure}

In order to verify this assumption, we perform an ``island counting''
measurement on the KPZ ensembles with the ultimate goal of deriving
the number of islands versus time statistics, $N(t)$. Islands are
interpreted in a straightforward way: an island is created in the
associated PNG system whenever a pinning event occurs, and two islands
adjoin to form one at the facet summits of the interface, reducing the
actual number of islands by one. An illustration of the island
construction is shown in Fig.~\ref{png_mapping}. This thought
experiment leaves us with a way to construct the associated island
number distribution in function of the time (height) in the KPZ
model. The $N(t)$ function for the individual samples is naturally
composed of step functions taking on the appropriate integer constants
between the respective time (height) intervals. The PNG and KPZ
behaviors are in good correspondence, which is a strong evidence for
the validity of our assumptions. The experimental relation for $N(t)$
calculated in KPZ systems is shown in Fig.~\ref{island_number}
together with the curve fitted based on the KAJM result,
Eq.~(\ref{n1}).

An associated quantity is the average height of the interface in time,
which can be measured in a very simple way in contrast to $N(t)$
mentioned in the preceding paragraphs. Let $S(t)$ denote the uncovered
fraction of a PNG system, Eq.~(\ref{n2}). Thus,
\begin{eqnarray}
\langle h_{\mathit{KPZ}}(t) \rangle = {1 \over L}
\int_0^L h_{\mathit{KPZ}}(x,t) \, dx = \int_0^t S(\tau) \, d\tau.
\end{eqnarray}

\begin{figure}
\begin{center}
\includegraphics*[width=\figwidth]{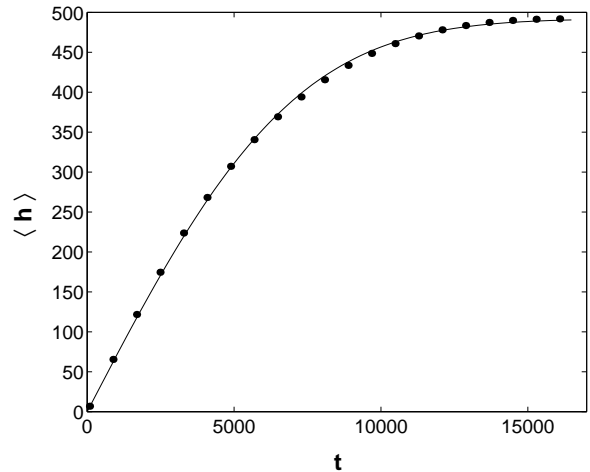}
\end{center}
\caption{The averaged interface height $\langle h \rangle$ versus time
$t$, the latter measured in lattice updating steps. A system of
$10000$ lattice units is used. The guide line is obtained from fitting
$\sqrt {\pi / 4} \, A \, {\mathrm{erf}}(B t)$ in a least-squares sense
with $A=555$ and $B=1.27 \cdot 10^{-4}$.}
\label{average_height}
\end{figure}

Fig.~\ref{average_height} shows the empirical relation found for
$\langle h_{\mathit{KPZ}}(t) \rangle$ and the adequately scaled error
function for a given interface evolution. Again, the theory proves to
be consistent with our experimental data to a high degree of
confidence.

\section{Summary}

In this paper we point out a mapping between interfaces in
two-dimensional systems, described by the negative-$\lambda$ quenched
Kardar--Parisi--Zhang equation and a one-dimensional island growth
model. An investigation of KPZ interfaces reveals that pinning
processes give rise to a morphology with valleys and facets that are
formed.  The facets are found to have straight edges with vanishing
deviations on a macroscopic scale, as well as sharp corners. This
establishes a connection with the PNG or KAJM model. The
correspondence is supported by evidence based on comparing statistical
quantities in both ensembles. Considering the KPZ model from this
point of view, it can be stated that pinning events are the
manifestation of a local nucleation process that is uncorrelated both
in time and space. The pinning probability for an interface segment
which advances covering an area of $dA$ only depends on $dA$ in an
exponentially decaying fashion.

The work presented here gives a geometric description of the
$\lambda<0$-depinning transition, in analogy to the directed
percolation depinning-picture of the opposite, $\lambda>0$-case. There
it is possible to extend the correspondence of the geometrical
construction and the pinned interfaces at $F_c$ to higher dimensions
\cite{resdio}. In our case this still awaits further work, but one
should note that the KAJM model of island growth becomes much more
difficult to handle, analytically, in $d>1$ \cite{seki,2dkajm}.

This work has been supported by the Academy
of Finland's Centre of Excellence Programme.


\end{multicols}

\bigskip

\begin{center}
\begin{minipage}{0.9\textwidth}
\begin{table}
\begin{center}
\begin{minipage}{0.6\textwidth}
\begin{tabular}{c|r|r @{ $\pm$ } l|c|c}
$F$ & \multicolumn{1}{c|}{Size} & \multicolumn{2}{c|}{Pinnings} &
\multicolumn{1}{c|}{Gradient left} & \multicolumn{1}{c}{Gradient right} \\
\hline

13.5 &  500 & 10.63 & 3.02 & $1.483 \pm 0.087$ & $1.484 \pm 0.086$ \\ 

     & 1000 & 10.26 & 2.07 & $1.483 \pm 0.066$ & $1.484 \pm 0.067$ \\

     & 2500 &  9.93 & 1.24 & $1.482 \pm 0.039$ & $1.482 \pm 0.040$ \\
\hline

14.0 &  500 &  5.80 & 2.13 & $1.652 \pm 0.085$ & $1.654 \pm 0.085$ \\

     & 1000 &  5.65 & 1.46 & $1.652 \pm 0.069$ & $1.653 \pm 0.066$ \\

     & 2500 &  5.53 & 0.89 & $1.651 \pm 0.041$ & $1.650 \pm 0.041$ \\
\hline

14.5 &  500 &  3.22 & 1.42 & $1.798 \pm 0.093$ & $1.797 \pm 0.095$ \\

     & 1000 &  2.85 & 1.02 & $1.802 \pm 0.068$ & $1.802 \pm 0.076$ \\

     & 2500 &  2.79 & 0.63 & $1.800 \pm 0.041$ & $1.800 \pm 0.042$ \\
\hline

14.8 &  500 &  2.49 & 0.94 & $1.869 \pm 0.085$ & $1.869 \pm 0.093$ \\

     & 1000 &  1.86 & 0.80 & $1.878 \pm 0.073$ & $1.879 \pm 0.069$ \\

     & 2500 &  1.74 & 0.50 & $1.879 \pm 0.039$ & $1.877 \pm 0.039$ \\
\end{tabular}
\end{minipage}
\end{center}
\medskip
\caption{Comparison table of interface morphology for
close-to-critical systems with different external driving forces; $F_c
\approx 15$. The ``Pinnings'' column contains the average number of
valleys found in the systems, normalized to 1000 lattice units. The
``Gradient left'' and ``Gradient right'' columns show the average
slope on the left hand side and right hand side of the facets,
respectively. The number of interface samples averaged is 10000 in
each of the cases. The errors are measured by the standard deviation
of the respective quantities.}
\label{Tmorphology_res}
\end{table}
\end{minipage}
\end{center}

\end{document}